\documentclass[twocolumn,showpacs,preprintnumbers,amsmath,amssymb]{revtex4}

\usepackage{graphicx}
\usepackage{dcolumn}
\usepackage{bm}

\begin{document}

\title{Possible resolution of the Casimir force finite temperature correction ``controversies"}

\author{S. K. Lamoreaux}

\affiliation{Yale University, Department of Physics, P.O. Box 208120, New Haven, CT 06520-8120}

\date{Jan. 6, 2008}

\begin{abstract}

By considering the effect of diffusion on the external electric
field response of charge carriers in metals and semiconductors, it
is shown that the finite temperature correction proposed Bostr\"om
and Sernelius requires substantial modification, and there is no
large correction as suggested for good conductors.  The apparent violation of the Third
Law of Thermodynamics of the various proposed temperature
corrections to the Casimir force is also resolved. Finally, the
effect of Debye screening on electrostatic calibrations between pure
germanium surfaces is calculated.

\end{abstract}
 \maketitle

\section{Introduction}

In our recent experimental work aimed at a precision measurement of
the Casimir force between pure germanium (Ge) plates \cite{1}, we
have discovered that the electrostatic calibration for Ge, a
semiconductor, is substantially different from what was expected,
assuming Ge to be a good conductor.  The problem is that a static
electric field can propagate a finite distance into a semiconductor;
this distance is determined by the combined consideration of
diffusion and field driven electric currents, leading to an
effective field penetration length (Debye-H\"uckel length)
\begin{equation}
\lambda=\sqrt{\epsilon \epsilon_0 kT\over e^2 c_t}
\end{equation}
where $c_t=c_h+c_e$ is the total carrier concentration, which for an
intrinsic semiconductor,  $c_e=c_h$. For intrinsic Ge
$\lambda\approx 1 \ \mu$m,while for a good conductor, it is less
than 1 nm.  $\lambda$ is independent of the applied field so long as
the applied field $E$ times $\lambda$ is less than the thermal
energy, $k_bT$ where $k_b$ is Boltzmann's constant.  In this limit,
and at sufficiently low frequencies and wavenumbers, thermal
diffusion dominates the field penetration into the material.  A
sufficiently low frequency for Ge would be $v_c/\lambda\sim 10$ GHz,
where $v_c$ is a typical thermal velocity of a carrier.

An analysis of the electrostatic energy between parallel plates is
given in the Appendix below, as well as the effect of field
amplitude on $\lambda$.  This analysis is crucial toward our ongoing
experimental efforts, especially for the electrostatic calibration.

In light of this analysis, it has become clear that none of the
recent papers describing finite temperature effects on the Casimir
force have taken into account the thermal diffusion of carriers
(electrons and/or holes) in the treatment of the boundary value
problem.  As such, a comprehensive review of recent work will not be
presented here; only the work by Bostr\"om and Sernelius, that led
to the recent ``controversy," will be discussed\cite{2}.

\section{Calculation of the Thermal Correction}

To calculate the effects of finite temperature,the electromagnetic
mode photon excitation number of $1/2$ due to zero point
fluctuations is replaced by
\begin{equation}
n(\omega)=\coth\left[{\hbar \omega \over 2 k_bT}\right]
\end{equation}
which has simple poles at
\begin{equation}
\omega_n={2\pi i k_bT\over \hbar}
\end{equation}
where in the following discussion we take only integers $n\geq 0$.

Following \cite{2}, the integral over $\omega$ in determining the
field energy between two flat plates is replaced by the sum over the
poles that occur at the Matsubara frequencies, $\omega_n$,
\begin{equation}
{\cal E}=k_bT {\sum_{n=0}^\infty}' f(\omega_n)
\end{equation}
where the prime indicates a factor of $1/2$ for the $n=0$ term, and
for the $TE$ modes (electric field parallel to the surface)
\begin{equation}
f(\omega_n)={1\over 2\pi}\int \ln \left[G^{TE}(q,i\omega_n)\right] q
dq
\end{equation}
where $q$ represents the electromagnetic field wavenumber in the
space between the plates, in direction perpendicular to the plates.

The function in the integral, for the $TE$ mode, is
\begin{equation}
G^{TE}(\omega_n,q)= 1-\left({\gamma_1-\gamma_0\over
\gamma_1+\gamma_0}\right)^2 e^{-2\gamma_0 d}
\end{equation}
\begin{equation}\label{modeq}
\gamma_i=\sqrt{q^2+\epsilon_i \omega ^2/c^2}
\end{equation}
where $i=0,1$ represents the space between the plates (0) or inside
the plates, (1), and $\epsilon_i$ is the respective electric
permittivity along the imaginary frequency axis, and $d$ is the
plate separation.

It is argued in \cite{2} that for realistic materials,
$G^{TE}(0,q)=1$ and hence does not contribute to the energy that
leads to the Casimir force between the plates.  For example, the
low-frequency permittivity of a metal is given by the conductivity
$\sigma$,
\begin{equation}
\epsilon_1(i \omega)= {4 \pi \sigma\over \omega}
\end{equation}
for which $\gamma_1=\gamma_2$ at $\omega=0$.For distances greater
than a few $\mu$ m, for $T=300$ K, the net force is reduced by a
factor of two compared to what is expected for a near perfect
conductor if both the $TE$ and $TM$ modes are included.

This result is in contradiction to experimental results,
particularly \cite{3}. There has been much discussion of this
correction, but to now, the effect of diffusive field screening has
not be taken into account in a satisfactory manner, or at all.

\section{Inclusion of the Debye Screening Length}

For realistic conducting materials (metal, semiconductor), Eq.
(\ref{modeq}) is not possibly correct, for we know an electric field
near the surface causes charges to move, and this tends to screen
out the applied field.  Electric fields applied either parallel or
perpendicular to the surface will be screened, varying compared to
the field $E_0$ at the surface, in the material, as
\begin{equation}
E(x)=E_0 e^{-|x|/\lambda}
\end{equation}
which is valid when$E_0\lambda<k_bT$ (or
$E_0\lambda<k_bT/\epsilon_1$ for fields perpendicular to the
surface).

The fact that electric fields do not penetrate any appreciable
distance into even very poor conductors is experimentally
well-known. Eq. (\ref{modeq}) incorrectly describes the observed
penetration of low frequency fields into conductors.

The screening effect can be treated in a heuristic fashion.  The
exact solution to the combined electrodynamic and diffusion problem
is beyond the scope of the present note, the intent of which is to
illustrate an effect that has been overlooked. Noting that $\lambda$
is independent of frequency at low frequencies ($<10^{10}$ Hz), and
that $E$ and its derivative are continuous across the boundary, we
can rewrite Eq. {\ref{modeq}) to include the spatial variation due
to Debye screening,
\begin{equation}\label{modeq2}
\gamma_1=\sqrt{q^2+\lambda^{-2}+\epsilon_i \omega ^2/c^2}
\end{equation}
assures continuity of the derivative across the boundary.  The point
is that the wavenumber in the material cannot be less than
$1/\lambda$.

For this modified function, for $\omega\rightarrow 0$, $G^{TE}$ is
equivalent to the perfect conductor result.  Thus the large
correction suggested in \cite{2} applies only to materials where
charges are not free to move, and diffusive effects do not enter. We
can question whether this condition can ever really be met, but for
any realistic slightly conducting material at finite temperature,
there will always be a finite screening length, and hence a full
contribution from the $G^{TE}$, $n=0$ mode, for good conductors 
where $\lambda$ is large to frequencies of order $10^{13}$ Hz.  A detailed
calculation for Ge is required because $\lambda$ starts falling off for frequencies
above 10 GHz, in the region where the thermally excited photons contribute most
significantly \cite{4}.

\section{Violation of the Third Law?}

It has been suggested that the $\omega=0$ term in Eq. () show a
manifest violation of the Third Law of Thermodynamics (sometimes
referred to as the Nernst Heat Theorem) because the system entropy
(identifying $\cal E$ as the free energy $F$), given by
\begin{equation}
S={\partial F\over \partial T}\big\vert_V=k_b f(0)/2
\end{equation}
is not zero in the limit $T\rightarrow 0$.This conundrum has been
addressed and clarified in \cite{5}, but there is a simpler argument
that will be presented here.

In particular, we can question what it means to convert the integral
over $\omega$ into a contour integral and hence sum over the
Matsurba frequencies.  If we did not convert this integral into a
sum, the separate identification of the zero frequency contribution
would not be made. Specifically, $T$ does not appear in the Casimir
force calculation expect through the mode photon number.
Differentiating Eq. (2) with $T$, we find
\begin{equation}
{d n(\omega)\over d T} =-\left[{1\over e^{\hbar\omega/
k_bT}-1}\right]^2\left[{\hbar \omega \over k_bT^2} e^{\hbar\omega/
k_bT}\right]
\end{equation}
which indeed goes to zero at $T=0$.

Alternatively, if we consider the entire sum, Eq. (4), let $T$ go to
zero, we find,
\begin{equation}
{\cal E}=k_bT {\sum_{n=0}^\infty} ' f(\omega_n)=k_b T\int f(2\pi  n k_bT/ \hbar ) dn
=  {\hbar\over 2 \pi}\int f(\omega) d \omega
\end{equation}
where the simple substitution $2\pi n k_b T/\hbar =\omega$ was made.
Hence, the temperature does not appear explicitly in the total free
energy, and the entropy indeed goes to zero at zero temperature.

The apparent violation of the Third Law is due simply to the
isolation of a single term in the total free energy.  Not
considering the entire system in calculating the entropy is
generally considered a sophomoric error.

\section{Conclusion}

By including the effect of charge movement and screening through the
Debye length, it is shown that the large correction to the Casimir
force predicted in \cite{2} is not applicable to realistic
materials.  It should be noted that these correction apply to all
conductors when the distance scale approaches the Debye length,
which for a good conductor is 0.1 nm. The Debye length is constant in good
conductors up to frequencies of order $10^{14}$ Hz, so we can expect the
full perfectly conducting force for any metal, at large separations.

The case of a semiconductor like Ge is slightly more
complicated because $\lambda$ begins to increase for frequencies of order
10 GHz, so a detailed anlysis of this case is required.  However, it
can be expected that there is a significant contribution from the $TE,\ n=0$ mode.

It is also shown that the apparent violation of the 3rd Law of
Thermodynamics is due to the isolation of a single term in an
expansion that becomes an integral in the limit of $T\rightarrow 0$.

\section{Appendix}

The potential in a plane semiconductor, if the potential is defined
on a surface $x=0$ is
\begin{equation}
V(x)=V(0)e^{-|x|/\lambda}
\end{equation}
where $\lambda$ is the
Debye-H\"uckle screening length, defined previously.

We are interested in finding the energy between two thick Ge plates
separated by a distance $d$, with a voltages $+V/2$ and $-V/2$
applied to the backs of the plates.  In this case, the field is
normal to the surface. After we find the energy per unit area, we
can use the Proximity Force Theorem to get the attractive force
between a spherical and flat plate.

Let $x=0$ refer to the surface of the plate 1, and $x=d$ refer to
the surface of plate 2.  By symmetry, the potential at the center
position between the plates is zero.  The potential in plate 1 can
be written
\begin{equation}
V_1(x)=V/2-(V/2-V_s) e^{-|x|/\lambda}
\end{equation}
and the space between the plates
$$V_0(x)=-2 V_s x/d +V_s$$
where we assume the field is uniform.  $V_s$, the surface potential,
is to be determined.

We need only consider the boundary conditions in plate 1, which are
$$V_1(-\infty)=V/2$$
$$V_0(0)=V_1(0)$$
(which has already been used)
$$\epsilon {d V_1(x)\over dx}\vert_{x=0}={d V_0(x)\over dx}\vert_{x=0}$$
where the last two imply that $D=\epsilon E$ is continuous across
the boundary.

The solution is
\begin{equation}
V_s={V\over 2}\left({1\over 1+2\lambda/\epsilon d}\right).
\end{equation}
With this result, it is straightforward to calculate the total field
energy per unit area in both plates and in the space between the
plates. The result is
\begin{equation}
{\cal E}={1\over 2} {\epsilon_0 V^2\over d}\left[{y+y^2\over
(y+2)^2}\right]
\end{equation}
where the dimensionless length $y=\epsilon d/\lambda$ has been
introduced.

If $V-V_s$ is large compared to $k_bT$, the effective penetration
depth increases because the charge density is modified in the
vicinity of the surface.  The potential in the plates is no longer a
simple exponential, however one can define an effective shielding
length \cite{6}
\begin{equation}
{\lambda'\over \lambda}={|\Phi|\over \sqrt{e^{\Phi}+e^{-\Phi}-2}}
\end{equation}
where
\begin{equation}
\Phi={V/2-V_s\over k_bT}
\end{equation}
with the results plotted in Fig. 2.  Given that $k_bT=30$ meV, at
plate separations of order 1 $\mu$m for Ge this begins to be a large
correction when voltages larger than 60 mV are applied between the
plates at separations of order $1\ \mu$m.

\begin{figure}
\begin{center}
\includegraphics[
width=4in ] {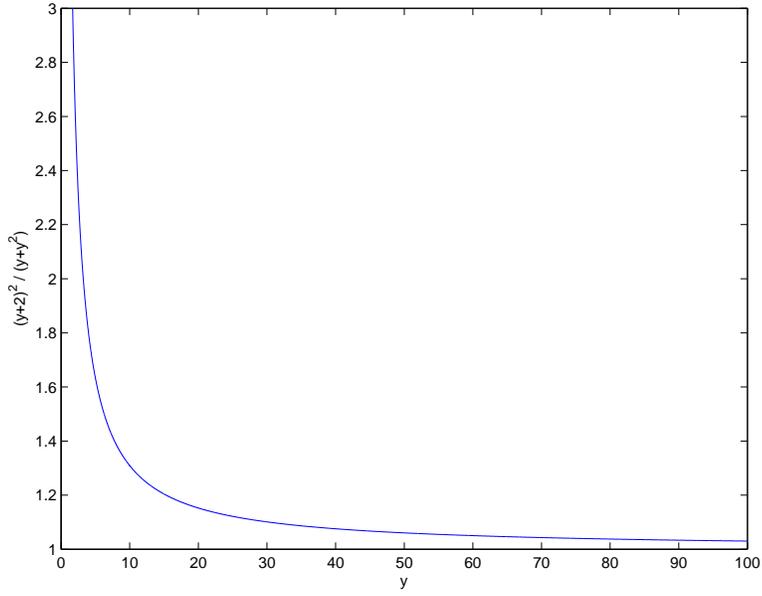} \caption{The separation determined by the
force resulting from an applied potential, assuming a $1/d$ force
variation, should be multiplied by this value.  For Ge, $\epsilon
d/\lambda=20$ corresponds to 1 $\mu$m plate separation.}
\end{center}
\end{figure}

\begin{figure}
\begin{center}
\includegraphics[
width=4in ] {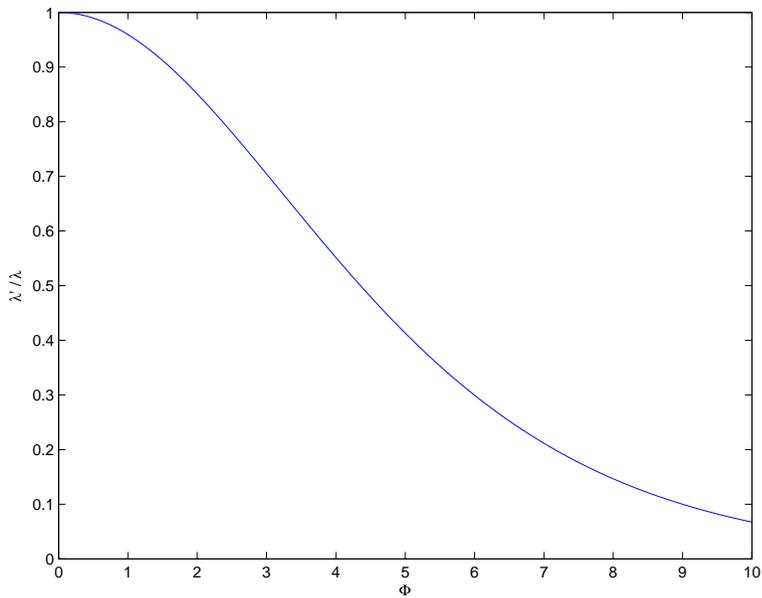} \caption{Effective field propagation
distance as a function of the dimensionless potential difference between the bulk
and surface.}
\end{center}
\end{figure}


\begin{thebibliography}{99}

\bibitem{1} S.K. Lamoreaux and W.T. Buttler, Phys. Rev. E {\bf 71},
036109 (2005).

\bibitem{2} M. Bostr\"om and B. Sernlius, Phys. Rev. Lett. {\bf 84},
4757 (2000).

\bibitem{3} A. Lambrecht, S. Reynaud, and S.K. Lamoreaux, Phys. Rev.
Lett. {\bf 84}, 5672 (2000), and references therein.

\bibitem{4} J.R. Torgerson and S.K. Lamoreaux, Phys. Rev. E {\bf 70}, 47102 (2004).

\bibitem{5} M. Bostr\"om and B.E. Sernelius, Physica A {\bf 339}, 53
(2004).

\bibitem{6} A.I. Spitsyn and V.M. Vanstan, Radiophysics and Quantum
Electronics {\bf 36}, 752 (1994).


\end{thebibliography}
\end{document}